\documentstyle[preprint,aps]{revtex}
\begin{document}
\bibliographystyle{plain}
\title{
Deducing correlation parameters from optical conductivity
in the Bechgaard salts
}
\vskip0.5truecm
\author {Fr\'ed\'eric Mila}
\vskip0.5truecm
\address{
      Laboratoire de Physique Quantique, Universit\'e Paul Sabatier\\
      31062 Toulouse (France)\\
      }
\maketitle

\begin{abstract}
Numerical calculations of the kinetic energy of various extensions of
the one-dimensional Hubbard model including dimerization and repulsion between
nearest neighbours are reported.
Using the sum rule that relates the kinetic energy to the integral of the
optical conductivity, one can determine which parameters
are consistent with the reduction
of the infrared oscillator strength that has been observed in
the Bechgaard salts. This leads to improved estimates of
the correlation parameters for both the TMTSF and TMTTF series.
\end{abstract}
\vskip .1truein

\newpage


The problem of finding an accurate description of the electronic properties of
quasi-one-dimensional organic conductors has a long history. Even for the
one-dimensional properties, which can be observed at not too low temperatures,
there is no consensus. One of the origins of the difficulty
is that one can work within two different kinds of model. The first one is the
Fermi gas model also known as $g$-ology\cite{solyom,emery}. This is the
appropriate framework to
describe the low-energy properties which are not of the usual Fermi liquid type
but of the Luttinger liquid type. Most of the low-energy properties have been
extensively analyzed within this type of model, and some information on the
size
of the parameters is available\cite{bourbonnais,wzietek}. The second
kind of model is the Hubbard model and its extensions. They are described by
the
Hamiltonian
\begin{eqnarray}
H = -t_1 \sum_{i {\rm even}, \sigma} (c_{i,\sigma}^\dagger
c_{i+1,\sigma}^{\vphantom{\dagger}} + {\rm h.c.})
      -t_2 \sum_{i {\rm odd}, \sigma} (c_{i,\sigma}^\dagger
      c_{i+1,\sigma}^{\vphantom{\dagger}} + {\rm h.c.})
      + U \sum_{i} n_{i\uparrow}n_{i\downarrow}
      +V  \sum_{i} n_{i}  n_{i+1}
\end{eqnarray}
The parameters of this model are: i) a hopping integral $t_1$ for the short
bonds; ii) a hopping integral $t_2$ ($\le t_1$) for the long bonds; iii) an
on-site repulsion $U$; iv) a repulsion between nearest neighbours $V$.
In the following, energies will be measured in units of
$t_1$, and the basic dimensionless parameters are $t_2/t_1$ for the
dimerization
and $U/t_1$ and $V/t_1$ for the Coulomb interactions. These parameters are
large
energy scales, and such a description is the natural framework to make contact
with quantum chemistry or to analyze high-energy properties of the materials.

Recently, it has been shown that, even
for the low-energy properties, this description in terms of high energy scales
can be very
useful\cite{milazotos}.
In the context of the Luttinger liquid theory, the
central parameter that describes the low-energy physics, the exponent $K_\rho$,
can take any value from $0$ to $+ \infty$. However, for the
quarter-filled, extended Hubbard model, which is the appropriate description of
(TMTSF)$_2$PF$_6$ as far
as quantum chemistry is concerned, it was possible to show that $K_\rho$ cannot
be smaller than $1/4$. It is actually  possible to calculate $K_\rho$ as a
function of the correlation parameters with a
reasonable accuracy using standard
numerical techniques\cite{schulz,ogata,milazotos}, and an accurate
determination
of these parameters will be an important step toward a good understanding of
the low-energy properties of the Bechgaard salts.

How can one determine the correlation parameters? In systems like
transition metal compounds, the combination of photoemission and Auger
spectroscopy has proved to be the most powerful tool to answer that
question\cite{sawatzky},
but this is hopeless in the case of the Bechgaard salts because the
photoemission spectrum is itself the subject of much
controversy\cite{dardel,milazotos}.
An analysis of the charge-transfer spectra of many charge-transfer salts
by Mazumdar and
Dixit led them to the general conclusion
that $U$ is about 1.5 eV and $V$ is 0.5 eV\cite{mazumdar}.
This analysis is certainly relevant, but in view of the very numerous
experimental
results that are now available for the Bechgaard salts, an analysis
more specific to these compounds, including in particular the effect of
dimerization, should be performed.
The
interpretation of the temperature dependence of the
susceptibility\cite{mila2,bourbonnais} and of the
resistivity\cite{milapenc} has given some preliminary information,
but it is not very precise:
Quantitative calculations are already very difficult for the Hubbard model with
only on-site repulsion, and accurate estimates for the model of Eq. (1) as a
function of  $t_2/t_1$, $U/t_1$ and  $V/t_1$ are not available. Finally,
quantum
chemistry calculations of the correlation parameters have also been performed
for these systems\cite{fritsch}, but the reliability of the
calculation is again very controversial.

There is however one set of
experimental data that has not been systematically used,
namely the reflectivity measurements. That these results contain information on
the local correlations has already been explained in great details by
Jacobsen\cite{jacobsen}.
The
idea is the following. On one hand, one can determine the plasma frequency from
a Drude fit
of the reflectivity spectrum. On the other hand, one can extract the optical
conductivity through
a standard procedure. Comparing the integral of the real part of the
conductivity with the plasma frequency yields an estimate of the reduction of
the kinetic energy due to correlations. To be more precise, let us denote by
$T$
the kinetic energy operator defined by
\begin{eqnarray}
T = -t_1 \sum_{i {\rm even}, \sigma} (c_{i,\sigma}^\dagger
c_{i+1,\sigma}^{\vphantom{\dagger}} + {\rm h.c.})
      -t_2 \sum_{i {\rm odd}, \sigma} (c_{i,\sigma}^\dagger
      c_{i+1,\sigma}^{\vphantom{\dagger}} + {\rm h.c.})
\end{eqnarray}
The kinetic energy is given by $E_{\rm kin}=<T>$, where the expectation
value is
calculated in the ground-state of the full Hamiltonian $H$. The kinetic energy
per site ${\cal E}_{\rm kin}$ is then defined as $\lim_{L\rightarrow +\infty}
E_{\rm kin}/L$, where $L$ is the number of sites. The plasma frequency
 provides an
estimate of the kinetic energy ${\cal E}_{\rm kin}^0$ calculated in the
ground-state
of $T$, i.e. without correlations, the integral of the conductivity
provides an estimate of ${\cal E}_{\rm kin}$, and the reduction of kinetic
energy is
defined as the ratio ${\cal E}_{\rm kin}/{\cal E}_{\rm kin}^0$.
The main difficulty is where to stop in
performing the integral of the conductivity. This can be a serious problem
because, for strongly correlated systems, spectral weight coming from the
conduction band can be found at high energy - typically around $U$ - and it is
impossible to disentangle this weight from other contributions to the
conductivity. In the case of the Bechgaard salts, this is not too serious
because
the systems are essentially quarter-filled. In that case, it was shown by
Maldague\cite{maldague}, and recently confirmed by Eskes and Oles\cite{eskes},
that the sum rule is almost
exhausted by the lower band, and the estimates of the kinetic energy obtained
by
integrating over the lower band only are accurate to within a few percents.

Performing such an analysis for the Bechgaard salts, Jacobsen reached the
conclusion that    ${\cal E}_{\rm kin}/{\cal E}_{\rm kin}^0$ is about
0.85 for (TMTSF)$_2$ClO$_4$ and
0.73 for (TMTTF)$_2$PF$_6$. Various estimates of the hopping
integrals for (TMTTF)$_2$PF$_6$ and (TMTSF)$_2$ClO$_4$ have been proposed on
the
basis of experimental results and quantum chemistry calculations. While there
is still
some uncertainty concerning their absolute value, especially in the case of
(TMTTF)$_2$PF$_6$, the ratio $t_2/t_1$ is believed to be approximately equal to
0.9 for (TMTSF)$_2$ClO$_4$ and 0.7 for (TMTTF)$_2$PF$_6$\cite{ducasse}.
So to extract
information about the correlation parameters from Jacobsen measurements, one
just needs accurate estimates of the kinetic energy in the ground-state of the
Hamiltonian of Eq. (1) as a function of $U/t_1$ and  $V/t_1$.
When Jacobsen
published his results, nothing of the sort was available, and he could not go
beyond a qualitative analysis of the results based on the numerical study of
a system of 2 particles on 4 sites. Motivated by Jacobsen's results, Baeriswyl
et al\cite{baeriswyl} calculated the kinetic energy of the standard
Hubbard model using the
Bethe
ansatz solution at half-filling and the Gutzwiller ansatz away from
half-filling. Their results confirm the trends, namely that correlations have
to
be invoked to explain the reduction of oscillator strength, but they do not
allow for a precise interpretation of the experimental results. We are not
aware
of any further work on that problem.

In this paper, we calculate the kinetic energy
finite clusters
on the basis of
numerical results
obtained on finite clusters by exact diagonalization. For a given cluster, the
expectation value of the kinetic energy is most easily obtained using the
Hellmann-Feynman theorem as
\begin{equation}
E_{\rm kin}=t_1\ \partial_{t_1}E_{G.S.} + t_2\ \partial_{t_2}E_{G.S.}
\end{equation}
where the ground-state energy $E_{G.S.}\equiv <H>$ is evaluated with Lanczos
algorithm.
${\cal E}_{\rm kin}$ is then obtained from
a finite-size scaling
analysis of $E_{\rm kin}/L$.
For non-interacting electrons, it is easy to show that the finite-size
corrections  go as $1/L^2$, where $L$ is the number of sites. This remains
true for the total energy per site for
Luttinger liquids. So it is quite natural to assume that this is also true for
the derivatives of this quantity with respect to the hopping integrals, and
thus
for the kinetic energy. Our numerical results clearly support this assumption.
A typical example is given in Fig. (1), where we have plotted
$E_{\rm kin}/L$ as a function of $1/L^2$ for $L=8$, $12$ and $16$. The two
curves
have been obtained for different
boundary conditions correponding to closed  and open shells respectively.
The scaling law is quite accurately satisfied,
and the two boundary conditions give estimates in very good agreement, which
lends further support in favour of the assumption that the scaling is in
$1/L^2$. In most
cases where we have tried both types of boundary conditions, the slope was
smaller for the boundary conditions corresponding to open shells, and all the
results given in the rest of the paper
have been obtained with such boundary conditions. Besides, we have compared the
results obtained by using only $L=8$ and $L=12$ with estimates obtained using
also the results for $L=16$ for a few cases, and the error was always less
than 1\%. So, unless one needs a very accurate value of the kinetic energy,
it is sufficcient to use systems with 8 and 12 sites to perform the $1/L^2$
extrapolation. This has been systematically done in the following.


Let us start with the results obtained in
the case $V/t_1=0$ (no repulsion between first neighbours).
The ratio  ${\cal E}_{\rm kin}/{\cal E}_{\rm kin}^0$is
plotted as a function of $U/t_1$ for three values of $t_2/t_1$ in Fig. (2) for
a
quarter-filled system. In all
the cases, it decreases with $U$, in agreement with the intuitive idea that
correlations make the motion more difficult and lead to a reduction of kinetic
energy. What is maybe more surprising is that, even for very large values of
$U/t_1$,
the reduction is not so big. That our results are still valid for large $U/t_1$
can actually be
checked quite easily. The value for $U/t_1 \rightarrow +\infty$ is the same as
for spinless fermions at half-filling (see Table I), and our results plotted as
a function of $t_1/U$ extrapolate nicely toward that liniting value
(see Fig. (3)).
So, taking for granted that $t_2/t_1=.7$
for (TMTTF)$_2$PF$_6$ , a reduction of 0.73 is incompatible with on-site
repulsion only.
This can be considered as a proof that repulsion on
neighbouring sites is important in these systems.


Let us now consider the general case described by Eq. (1). The question we
would like to answer is
the following: What are the values of $U/t_1$ and $V/t_1$ that are compatible
with the known values of $t_2/t_1$ and of
${\cal E}_{\rm kin}/{\cal E}_{\rm kin}^0$
for (TMTSF)$_2$ClO$_4$ and
(TMTTF)$_2$PF$_6$? The most convenient thing to do is to plot the curves of
constant ${\cal E}_{\rm kin}/{\cal E}_{\rm kin}^0$
in the ($U/t_1$, $V/t_1$) plane
for the values of
$t_2/t_1$ of interest. Such plots for quarter-filled systems
are given in Fig. (4) for $t_2/t_1=1$, $0.9$
and $0.7$. The basic features of these curves are again quite natural. The only
one that deserves a special comment is the reentrant behaviour for large
$V/t_1$
and small $U/t_1$. Another way of looking at the same effect is to notice that,
for a given value of $V/t_1$, the kinetic energy first increases before it
decreases for $U/t_1$ large enough. This presumably comes from local pairs,
which are known to exist and to be very heavy objects in the small $U/t_1$,
large $V/t_1$ limit\cite{penc2}, and which become lighter when $U/t_1$
increases.

There is now no problem to find parameters that give a reduction of kinetic
energy of 0.85 for (TMTSF)$_2$ClO$_4$ and of 0.73 for (TMTTF)$_2$PF$_6$. The
corresponding curves of possible pairs ($U/t_1$, $V/t_1$) are reproduced in
Fig. (5). This is not very useful unless we can decide where the actual
parameters are located on these curves. This is actually possible on the basis
of general arguments coming from quantum chemistry. Unlike the $g$-parameters
of the Fermi-gas model, $U$ and $V$ have a simple microscopic meaning: $U$ is
the energy needed to put two particles on the same site, and $V$ is the energy
needed to put them on neighbouring sites. Now, the molecules TMTSF and TMTTF
are
very similar, the only difference being that the $3p$-orbitals of sulfur
in TMTTF
are more concentrated than the $4p$-orbitals of selenium in TMTSF, and quantum
chemistry calculations predict that the ratio $U({\rm TMTTF})/U({\rm TMTSF})$
is in the range
1.0 - 1.25. We also know from different sources\cite{jacobsen,ducasse} that
$t_1({\rm TMTTF})/t_1({\rm TMTSF})$ is in the range 0.7 - 0.8.
So the ratio $U/t_1$ is at
most 80\% larger for TMTTF than for TMTSF. But from Fig. (5), we know that this
ratio is at most 8 for TMTSF, so it is at most 15 for TMTTF. Looking again at
Fig. (5), this means that $V/t_1$ is at least equal to 2 for TMTTF.
Now, on the basis of
quantum chemistry, $V$ is expected to be roughly the same for both types of
compounds, which implies that $V/t_1$ is at least 1.5 for TMTSF. But according
to Fig. (5), this means that $U/t_1$ cannot be larger than 6 in that compound.
This again puts a constraint on $U/t_1$ in TMTTF, and so on. Finally, if we use
the constraints given by quantum chemistry, we end up with the
parameters given in Table 2. These values are just estimates, and one should
put
error bars on them.
The main
source of uncertainty probably comes from the
experimental results. In particular, the saturation value of the integral of
the
loss function generally gives numbers in good agreement with the plasma
frequency deduced from a Drude fit of the reflectivity spectrum.  However,
in the
present case, the values obtained from the loss function are slightly
larger\cite{jacobsen},
leading to smaller values of  ${\cal E}_{\rm kin}/{\cal E}_{\rm kin}^0$ (0.67
for (TMTTF)$_2$PF$_6$, 0.80 for (TMTSF)$_2$ClO$_4$). Another source  of
uncertainty lies in the values used for the ratios $t_2/t_1$, but the
dependence of ${\cal E}_{\rm kin}/{\cal E}_{\rm kin}^0$ on this parameter is
smooth (see Fig. 4). Finally, the location on the curves is only approximative
because the arguments derived from quantum chemistry are only qualitative or,
at best,
semi-quantitative.

In spite of these sources of uncertainty, two points seem to be clearly
established. First, the reduction of oscillator strength reported for
TMTTF implies that the repulsion between first neighbours is not
negligible. Second,
$U/t_1$ is larger in TMTTF than in TMTSF. If that
was not the case, $V$ would have to be much larger in TMTTF than in TMTSF,
which
can be rejected on the basis of quantum chemistry. This last conclusion should
be contrasted with the interpretation of the temperature dependence of the
susceptibility of Wzietek et al\cite{wzietek}, which lead them to conclude
that $U/t_1$ is
about the same in both series. Recent calculations of the
susceptibility\cite{mila2}
suggest however that the temperature dependence is actually consistent with our
present conclusion that $U/t_1$ is larger in TMTTF than in TMTSF.

In conclusion, the reduction of oscillator strength reported for the Bechgaard
salts by Jacobsen leads to precise and useful information on the size of the
correlation parameters in these compounds. The main difference with respect to
the microscopic models used so far in the interpretation of various
experimental
results (susceptibility, minimum of resistivity,...) lies in the presence of a
relatively large value of the repulsion between nearest neighbours. Whether the
estimates proposed in the present paper are consistent with the other
experimental data remains to be seen.

I am very grateful to Jean-Paul Pouget, who encouraged me to look at the
optical
data of Jacobsen.

\begin{figure}
\caption{ Finite-size scaling of the kinetic energy for $t_2/t_1=1$,
$U/t_1=10$, and $V/t_1=0$. Upper curve: closed shell;
lower curve: open shell.}
\end{figure}

\begin{figure}
\caption{ Kinetic energy as a function of $U/t_1$ for $V/t_1=0$. a)
$t_2/t_1=1$; b)  $t_2/t_1=0.9$; c)  $t_2/t_1=0.7$ }
\end{figure}

\begin{figure}
\caption{Large $U$ behaviour of the kinetic energy for $V/t_1=0$. a)
$t_2/t_1=0.9$; b) $t_2/t_1=0.7$ }
\end{figure}

\begin{figure}
\caption{Constant kinetic energy plots from
${\cal E}_{\rm kin}/{\cal E}_{\rm kin}^0 =0.99$ (bottom left) to
${\cal E}_{\rm kin}/{\cal E}_{\rm kin}^0 =0.50$ (top right). a) $t_2/t_1=1$;
b)  $t_2/t_1=0.9$; c)
$t_2/t_1=0.7$}
\end{figure}

\begin{figure}
\caption{Comparison of the curves giving the correct reduction of kinetic
energy. Crosses:
(TMTSF)$_2$ClO$_4$ ($t_2/t_1=0.9$,
${\cal E}_{\rm kin}/{\cal E}_{\rm kin}^0 =0.85$); Squares:
(TMTTF)$_2$PF$_6$ ($t_2/t_1=0.7$,
${\cal E}_{\rm kin}/{\cal E}_{\rm kin}^0 =0.73$)}
\end{figure}

\begin{table*}
\begin{center}
\begin{tabular}{cccccccccccc}
$t_2/t_1$ & 1. & 0.9 & 0.8 & 0.7 & 0.6 & 0.5 & 0.4 & 0.3 & 0.2 & 0.1 & 0. \\
${\cal E}_{\rm kin}/{\cal E}_{\rm kin}^0$
&  .707  & .711 & .720 & .733 & .752 & .777 & .808 & .845 & .889 &
.940 & 1. \\
\end{tabular}
\caption{Reduction of kinetic energy for $U/t_1 \rightarrow + \infty$ and
$V/t_1=0$ }
\end{center}
\begin{center}
\begin{tabular}{lccc}
Compound & $t_2/t_1$ & $U/t_1$ & $V/t_1$ \\\hline
(TMTSF)$_2$ClO$_4$ & 0.9  & 5.0 & 2.0 \\
(TMTTF)$_2$PF$_6$  & 0.7 & 7.0 & 2.8 \\
\end{tabular}
\caption{}
\end{center}
\end{table*}

\end{document}